\documentclass
[preprint,11pt]{elsarticle}

\usepackage{amssymb}
\usepackage{graphics}
\usepackage{epsfig}
\usepackage{subfigure}
\usepackage{mathrsfs}
\usepackage{amsmath}
\usepackage{amsfonts}
\usepackage{amssymb}
\usepackage{graphicx}
\usepackage{longtable}
\usepackage{rotating}
\usepackage{multirow}

\journal{}

\begin{document}

\begin{frontmatter}




\title{\textbf{Holographic thermalization  in
noncommutative geometry}}

\renewcommand{\thefootnote}{\fnsymbol{footnote}}

\author{Xiao-Xiong Zeng $^{1}$\footnote {E-mail address: xxzengphysics@163.com }, Xian-Ming Liu $^{2}$\footnote {E-mail address: liuxianming1980@163.com}, Wen-Biao Liu (corresponding author) $^{3}$\footnote {E-mail address: wbliu@bnu.edu.cn}}
\address{$^1$School of Science, Chongqing Jiaotong University, Chongqing, 400074, China\

$^2$Department of Physics, Hubei University for Nationalities, Enshi, 445000,
Hubei, China\

$^3$Department of Physics, Institute of Theoretical Physics, Beijing Normal
University, Beijing, 100875, China}

\begin{abstract}
Gravitational collapse of a shell of dust  in noncommutative geometry is probed by the renormalized geodesic length,   which is dual to probe the thermalization by the   two-point correlation function  in the dual conformal field theory. We find  that larger the
 noncommutative parameter is,
  longer the thermalization time is, which implies that the large noncommutative parameter delays
   the thermalization process. We also  investigate how the  noncommutative parameter affects the thermalization velocity and thermalization acceleration.

\end{abstract}




\end{frontmatter}



\section{Introduction}

Gravity in noncommutative geometry\cite{Snyder, Seiberg} and noncommutative field theory \cite{Hashimoto, Maldacena}   has been investigated extensively in recent years. The reason for this  prevailing phenomenon maybe arises from a fact that the singularities in general relativity and ultraviolet divergences in
quantum field theory can be avoided in the noncommutative   framework.  Because in noncommutative geometry, coordinates in  a manifold  fail to commute in
analogy to the conventional noncommutativity among conjugate variables
in quantum mechanics,  which leads to a natural cut off due to the position uncertainty.

To study the properties of gravity in the noncommutative geometry,  it is important and necessary to find the black hole solutions in this background.  Because
spacetime as
a manifold of points breaks down at distance scale of the order of the Planck length, it was proposed  that the point-like object should be replaced by a smeared
object\cite{Smailagic1, Smailagic2}. In this case, the description mathematically by a Dirac-delta function distribution is substituted by  a Gaussian
distribution of minimal width $\sqrt{\theta}$, where $\theta$ is the smallest fundamental unit of an observable
area in the noncommutative coordinates.
The first noncommutative black hole solution was presented by Nicolini, Smailagic and Spallucci using the coordinate
coherent state method\cite{Nicolini}. It was found that
the curvature
singularity of the black hole is removed. Moreover, their method is consistent with
Lorentz invariance, unitarity and UV finiteness of quantum field theory, which  appears in the Weyl-Wigner-Moyal $\star-$product approach. Until now, there are many investigations on the properties of the noncommutative
black hole, such as thermodynamic properties\cite{thermodynamics}, quantized entropy and area of horizon\cite{spectrum}, quantum tunneling radiation\cite{radiation}, gravitational collapse solution\cite{solution}, strong gravitational lensing effect\cite{lens}, and so on. Especially recently, Hawking-Page phase transition\cite{Nicolini1108}, holographic entanglement entropy\cite{Fischler137}, and holographic superconductors\cite{Pramanik1401} have also been
investigated as the noncommutative
anti-de Sitter balck hole solution\cite{Mann2011} is given.

In this paper, we intend to investigate the non-equilibrium thermalization process in  noncommutative field theory from the viewpoint of holography.
Recent years, investigation on holographic thermalization has attracted more and more attentions of theoretical physicist.
The main
motivation maybe arises from a fact that
 the thermalization time of quark
gluon plasma produced in RHIC and LHC experiments predicted by the perturbation theory is longer than the experiment result \cite{Gyulassy}.
In order to investigate  the thermalization process, one should construct a
proper model in gravity \cite{Danielsson}. Now, there have been many models
 to study the   non-equilibrium thermalization process
  \cite{Garfinkle84, Garfnkle1202, Allais1201, Das343, Steineder, Wu1210, Gao, Buchel2013, Keranen2012, Craps1, Craps2,  Balasubramanian1, Balasubramanian2}. Among them, one elegant model is presented in \cite{Balasubramanian1, Balasubramanian2},  where  the two-point
correlation function, Wilson loop, and entanglement entropy were used to detect the thermalization.
Now, such an investigation has been  generalized to  many gravity models  \cite{GS,CK, Yang1, Zeng2013, Zeng2014,  Baron, Li3764, Baron1212, Arefeva, Hubeny,
 Arefeva6041, Balasubramanianeyal6066, Balasubramanian4, Balasubramanian3, Balasubramanian9, Cardoso, Hubeny2014, Pedraza}.

The purpose of this  paper is to investigate how the noncommutative parameter affects the thermalization process. In the dual conformal field theory,  we take the
two-point correlation  function as a thermalization probe \footnote{Expectation value of Wilson loop and entanglement entropy also can be treated as the thermalization probes, it has been found \cite{Balasubramanian1, Balasubramanian2} that all of them have similar behavior thus here we only use the two-point correlation.}
to study the
thermalization behavior. According to  the AdS/CFT correspondence,
 this process equals to probing the evolution of a shell that interpolates between a pure AdS and a noncommutative  AdS black brane by the geodesic.
 Concretely we first study the motion profile of the geodesic, and then the  renormalized geodesic length. For the spacetime with a horizon, we find for both the thermalization probes,  larger the
 noncommutative parameter is,
  longer the thermalization time is. For the spacetime without a horizon, we find the shell will not collapse all the time but will stop
in a  stable state.
 In addition, we also obtain the fitting functions of  the thermalization curve for both thermalization probes.  Based on the functions, we get  the thermalization
 velocity and thermalization acceleration.

The remainder of this paper is organized as follows. In the next section, we shall provide a brief review of the gravitational collapse solution in the noncommutative geometry. Then  in Section 3,  the  collapse of the shell is probed by making use of the  renormalized geodesic length. The last section is devoted to our   conclusions.

\section{The noncommutative  Vaidya AdS black branes}

In this section, we will give a brief review of the noncommutative Vaidya AdS black branes. For details, please see\cite{Mann2011}. As well known, the metric for a
static spherically symmetric noncommutative AdS black hole is\cite{Mann2011}
\begin{equation}
 ds^{2}=-f(r)dt^{2}+f^{-1}(r)dr^{2}+r^{2}d \phi ^{2}+r^{2}\sin^{2}\phi d\varphi^{2},\label{areacoorections1}
\end{equation}
where%
\begin{equation}
 f(r)=1-\frac{4 M \gamma(\frac{3}{2},\frac{r^2}{4\theta} )}{r \sqrt \pi}+\frac{r^2}{S^2},   \label{areacoorections2}
\end{equation}%
in which $S$ is the radius of the AdS, $M$ is  the total mass diffused throughout the region of linear size $\sqrt{\theta}$, $\theta$ comes from the noncommutator of $[x^{\mu},x^{\nu}]=i \theta^{\mu\nu}$ with  $\theta^{\mu\nu}=\theta  \text{diag}[\epsilon_1,\cdots, \epsilon_{D/2} ]$, and
$\gamma(\frac{3}{2},\frac{r^2}{4\theta})$ is the lower incomplete Gamma function defined by
\begin{equation}
\gamma(\frac{3}{2},\frac{r^2}{4\theta} )\equiv\int_0^{\frac{r^2}{4\theta}} t^{\frac{1}{2}} e^{-t} dt.
\end{equation}
The black hole temperature in the noncommutative geometry is given by
\begin{equation}
T_{NC}\equiv \frac{\kappa}{2 \pi}=\frac{1}{4 \pi} \frac{\partial f(r)}{\partial r}\mid_{r_h},
\end{equation}
where $r_h$ is the event horizon of the black hole determined by  $f(r_h)=0$.
In addition, according to the  properties  of gamma function
\begin{equation}
\gamma(a+1,x)=a \gamma(a,x)-x^a e^{-x},
\end{equation}
\begin{equation}
\gamma(\frac{1}{2},x^2)\equiv2 \int_0^x e^{-t^2} dt=\sqrt{\pi}\texttt{Erf}(x),
\end{equation}
 Eq.(\ref{areacoorections2}) changes into
 \begin{equation}
 f(r)=1-\frac{2M}{r}\texttt{Erf}(\frac{r}{2 \sqrt{\theta}})+\frac{r^2}{S^2}+\frac{2M}{\sqrt{\pi\theta}}e^{-\frac{r^2}{4\theta}},
\end{equation}
where $\texttt{Erf}(\frac{r}{2 \sqrt{\theta}})$ is a Gauss error function.
It is obvious that this black hole spacetime is closely dependent on the noncommutative parameter
$\theta$. As $\theta\rightarrow 0$, this background reduces to the
conventional Schwarzschild AdS black hole. In this case, noncommutative fluctuations are negligible and the spacetime can be well
described by a classical manifold.

 As done in \cite{Chamblin},
we can also consider the limit where the boundary of $AdS_{d+1}$ is $R^d$ instead of
$R\times S^{d+1}$, namely the so-called infinite volume limit. After  the coordinate
transformation $z=\frac{S^2}{r}$,  Eq.(\ref{areacoorections1}) and the components of metric in this case change into
\begin{equation}
 ds^{2}=\frac{1}{z^2}[-H(z)dt^{2}+H^{-1}(z)dz^{2}+dx_i^2],\label{metric}
\end{equation}
\begin{equation}
 H(z)=1-2M\texttt{Erf}(\frac{1}{2 \sqrt{\theta}z })z^3+\frac{2M z^2}{\sqrt{\pi \theta}}e^{-\frac{1}{4\theta z^2}},\label{h}
\end{equation}
where  $S$ has been set to one and $i=1,2$.
Introducing the
Eddington-Finkelstein coordinate system, namely
\begin{equation}
dv=dt-\frac{1}{H(z)}dz,
\end{equation}
the background spacetime in  Eq.(\ref{metric}) changes into
\begin{equation}
ds^2=\frac{1}{z^2} \left[ - H(z) d{v}^2 - 2 dz\ dv +
dx_i^2 \right]. \label{collpse}
\end{equation}
Now  noncommutative  Vaidya AdS black brane can be obtained by
freeing the mass parameter in Eq.(\ref{h}) as an arbitrary function of $v$. As stressed in \cite{Park}, in this case, the mass source includes
the new matter related to the noncommutativity as well as the matter on the shell. In other words, Eq.(\ref{collpse}) can be treated as the solution of the following field equation
\begin{equation}
R_{\mu\nu}-\frac{1}{2}Rg_{\mu\nu}+\Lambda g_{\mu\nu} = 8 \pi G (T_{\mu\nu}^{\theta}+T_{\mu\nu}^{m}),
\end{equation}
 where  $ T_{\mu\nu}^{\theta}$ is the energy-momentum tensor arising from the noncommutative background  and
\begin{equation}
T_{\mu\nu}^m\propto 2z^2\frac{dM(v)}{dv}\delta_{\mu v}\delta_{\nu v}.
\end{equation}
Here $M(v)$ is mass  of a collapsing noncommutative  black brane, which  is usually
chosen as the smooth function
\begin{equation}
M(v) = \frac{M}{2} \left( 1 + \tanh \frac{v}{v_0} \right),\label{M}
\end{equation}
where $v_0$ represents a finite shell thickness.
For  Eq.(\ref{M}),
in the limit $v\rightarrow-\infty$, the mass vanishes and the background in Eq.(\ref{collpse}) thus  corresponds to a pure  AdS space. In
the limit $v\rightarrow \infty$, the mass changes into a constant and so the background represents a static noncommutative   Schwarzschild  AdS black brane.

\section{Probe of the thermalization}
As the  model that describes the thermalization
process on the dual conformal field theory is constructed, we will choose the two-point correlation
function at equal time to explore how the noncommutative parameters affects
 the thermalization process.
According to the AdS/CFT correspondence,  the equal time two-point correlation function under the  the saddle-point
approximation can
be holographically approximated as \cite{Balasubramanian2, Balasubramanian61}
\begin{equation}
\langle {\cal{O}} (t_0,x_i) {\cal{O}}(t_0, x_i^{\prime})\rangle  \approx
e^{-\Delta { L}} ,\label{llll}
\end{equation}
if the conformal dimension $\Delta$ of scalar operator $\cal{O}$
is large enough, where
$L$ indicates the length of the bulk geodesic between the points $(t_0,
x_i)$ and $(t_0, x_i^{\prime})$ on the AdS boundary.  Usually the geodesic length above is divergent
 due to the contribution of the AdS boundary, one should eliminate the  divergent part and  use the renormalized geodesic length, defined by $\delta L=L+2\ln z_0$ \cite{Balasubramanian1, Balasubramanian2}, where   $z_0$ is a UV cut-off that can be read from the boundary conditions
\begin{equation}\label{regularization}
z(\frac{l}{2})=z_0, v(\frac{l}{2})=t_0,
\end{equation}
in which $l$ is the  boundary separation between the points lies entirely over the $x_1$ direction and $t_0$ is the time of the thermalization probe moving from the shell to the boundary, which will be called as thermalization time later. Next we would like to rename $x_1$ as $x$ and employ it to
    parameterize the trajectory such that the proper length is given by
 \begin{eqnarray}
L=2 \int_0^{\frac{l}{2}} dx \frac{\sqrt{\Pi}}{z} ,\label{false}
\end{eqnarray}
with
\begin{equation}
\Pi=1-2z'(x)v'(x) - H(v,z) v'(x)^2,
\end{equation}
To minimize the length of the geodesic, we need to solve the two equations of motion for $z(x)$  and  $v(x)$ respectively. Varying the length functional in Eq.(\ref{false}), we get
\begin{eqnarray} \label{gequation}
z(x)\sqrt{\Pi}\partial_x (\frac{z'(x)+H(v,z)v'(x)}{z(x)\sqrt{\Pi}})=\frac{1}{2}\frac{\partial H(v,z) }{\partial v(x)} v'^2(x),\nonumber\\
z(x)\sqrt{\Pi}\partial_x (\frac{v'(x)}{z(x)\sqrt{\Pi}})=\frac{1}{2}\frac{\partial H(v,z) }{\partial z(x)} v'^2(x)+ \frac{\Pi}{ z(x)}.
\end{eqnarray}
To solve these equations, we need
to consider the symmetry of the geodesic and impose the following initial conditions
\begin{equation}\label{initial}
z(0)=z_*,  v(0)=v_* , v'(0) =
z'(0) = 0.
\end{equation}

\begin{figure}
\centering
\subfigure[$$]{
\includegraphics[scale=0.55]{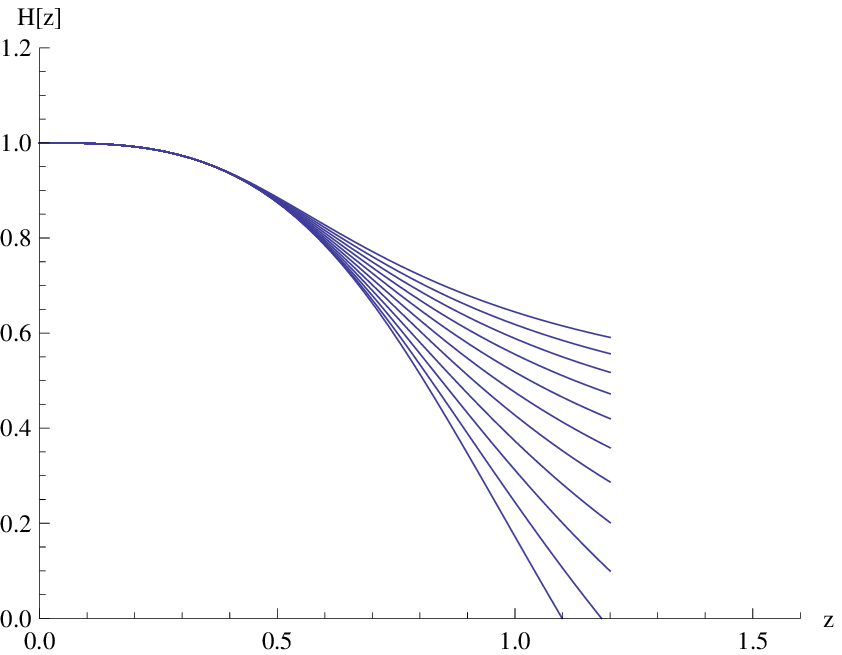}}
\subfigure[$$]{
\includegraphics[scale=0.45]{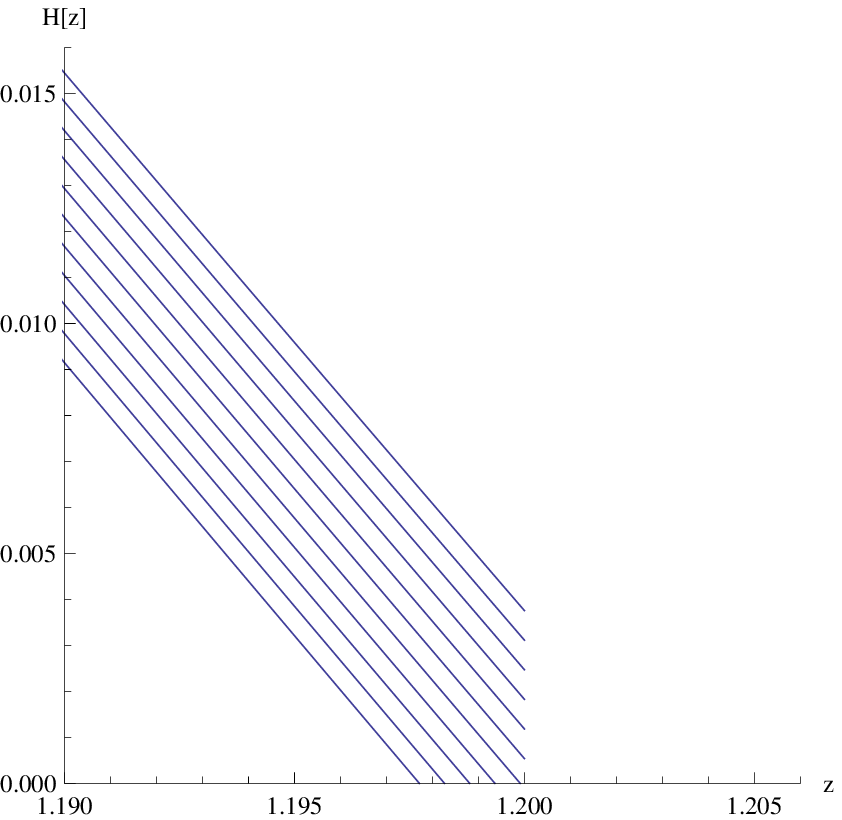}}
 \caption{\small In (a), $\theta$ changes from $0.1$ to $0.3$ with step  $\theta=0.02$, in (b), $\theta$ changes from $0.123$ to $0.124$ with step  $\theta=0.0001$.} \label{fig2}
\end{figure}

Next  we intend to solve the equations of motion  in  Eq.(\ref{gequation}) numerically with the help of the initial conditions in  Eq.(\ref{initial}).
 During the numerics, we will set the shell thickness $v_0 = 0.01$ and  UV
 cut-off $z_0 = 0.01$ respectively. In addition, the  mass $M$ will be set to $\frac{1}{2}$ as done in \cite{lens}.
In this case, with  Eq.(\ref{h}), we can check whether there is  a horizon for different noncommutative parameter $\theta$, which is plotted in  Figure (\ref{fig2}).  From  Figure (\ref{fig2}), we know that for  $\theta<0.1234$, there is always a horizon while for  $\theta>0.1234$,  there is not a horizon.
 That is, for  $\theta<0.1234$ a static black hole will be formed  at the last stage of the gravitational collapse process, which indicates that the non-equilibrium state will approach to an equilibrium state lastly from the viewpoint of duality.
For $\theta>0.1234$, though a black brane will not be formed, we will also use the  renormalized geodesic length to probe the  collapse of the shell so that we can know whether it  will collapse all the time. The time for the shell collapse from the pure  AdS  to the stable state is also called thermalization time though we can not define an equilibrium state strictly  in this case.

\begin{figure}
\centering
\subfigure[$\theta=0.01$]{
\includegraphics[scale=0.55]{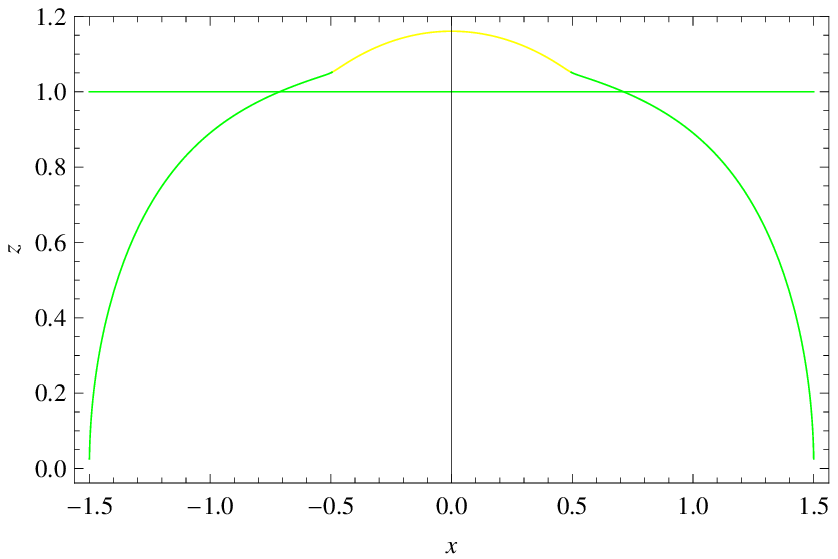}}
\subfigure[$\theta=0.1$]{
\includegraphics[scale=0.55]{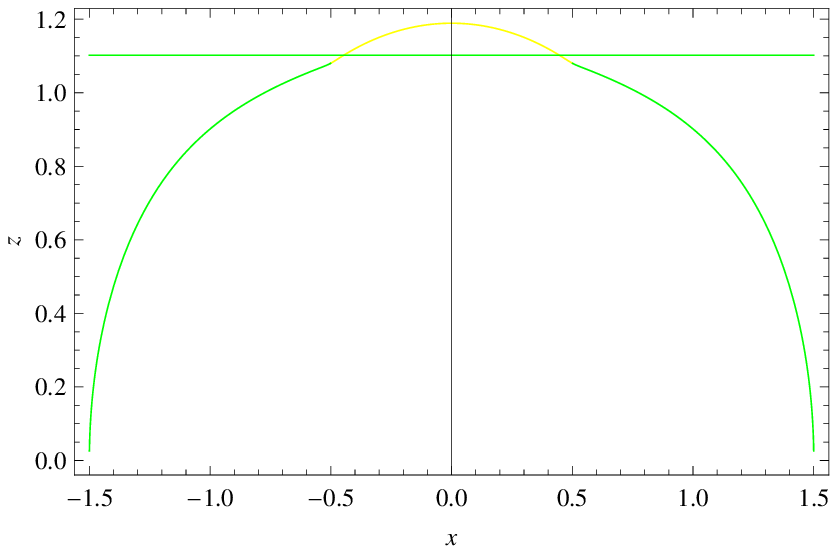}}
\subfigure[$\theta=0.3$]{
\includegraphics[scale=0.55]{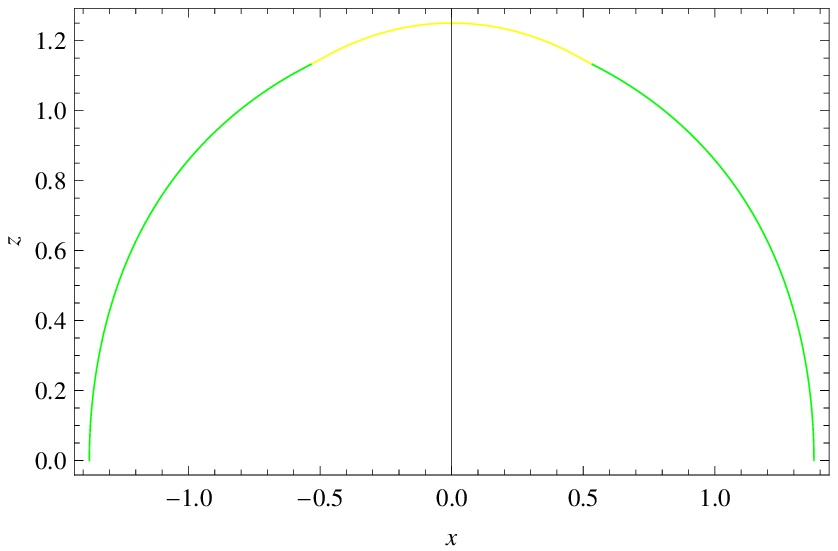}}
\subfigure[$\theta=0.5$]{
\includegraphics[scale=0.55]{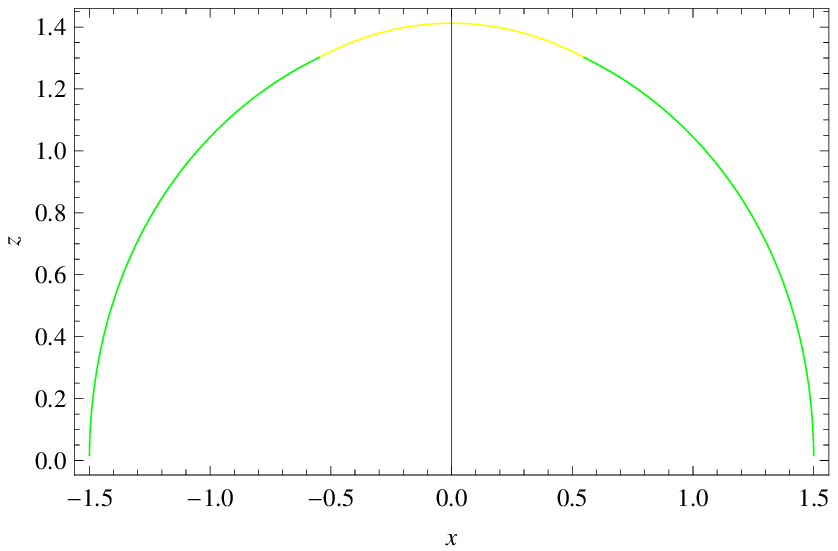}}
 \caption{\small Motion profile of the geodesic in the noncommutative  Vaidya AdS black brane with the same
boundary separation and initial time.  The black brane
horizon is indicated by the horizontal green line. The  position of the shell is described by the junction between the yellow line and green line.} \label{fig3}
\end{figure}

Firstly, we will set different initial time to explore whether the effect of the  noncommutative parameters is the same at different stage. In Table (\ref{tab1}), we list the thermalization time for different noncommutative parameters at different initial time  $v_*$. From it, we know that for a fixed initial time, as the noncommutative parameters raise, the thermalization  time increases firstly and then decreases step by step. Especially, for the large initial time, $v_*=-0.111$, the thermalization time decreases in advance. So we can conclude that the thermalization time for different noncommutative parameters is non-monotonic. In addition, at $v_*=-0.111$, we also plot the motion profiles of the geodesic
for different noncommutative parameters, which are shown in Figure (\ref{fig3}). In (a) and (b) in Figure (\ref{fig3}), we know that the spacetimes own horizons, thus we can distinguish whether a static black brane have been formed by checking whether the shell has been dropped into the horizon. It is obvious that the shell in   (a) has been dropped into the horizon while the shell in (b) is out of the horizon. A static black brane thus has been formed in   (a)  while the shell is collapsing in (b), which implies that the   quark
gluon plasma in the dual conformal theory  has been thermalized  for the case $\theta=0.01$ while it is thermalizing for the case $\theta=0.1$. In other words, as the noncommutative parameter increases, the thermalization will be delayed. In  (c) and (d) in Figure (\ref{fig3}), because there is not a horizon, we only know that the shell is collapsing, which implies  that the   quark
gluon plasma in the dual conformal theory   is thermalizing.

\begin{table}
\begin{center}\begin{tabular}{l|c|c|c|c|c}
 \hline
               &$\theta=0.01$ &  $\theta=0.1$  &$ \theta=0.3$   &  $\theta=0.5$  &   $\theta=0.7$    \\ \hline
$v_*$=-0.888    & 0.597428     &  0.597526      & 0.601945         & 0.605219          & 0.606315   \\ \hline
$v_*$=-0.444    &0.995173     &1.02105           &1.07147          &1.0729       & 1.06914     \\ \hline
$v_*$=-0.111     &1.27647      &1.33883        & 1.43042             &1.42512        &1.41624       \\ \hline
\end{tabular}
\end{center}
\caption{The thermalization time $t_0$ of the geodesic probe for different noncommutative parameters $\theta$ and different initial time $v_*$ with the same boundary separation $l=3$.}\label{tab1}
\end{table}

\begin{figure}
\centering
\subfigure[$l=3$]{
\includegraphics[scale=0.55]{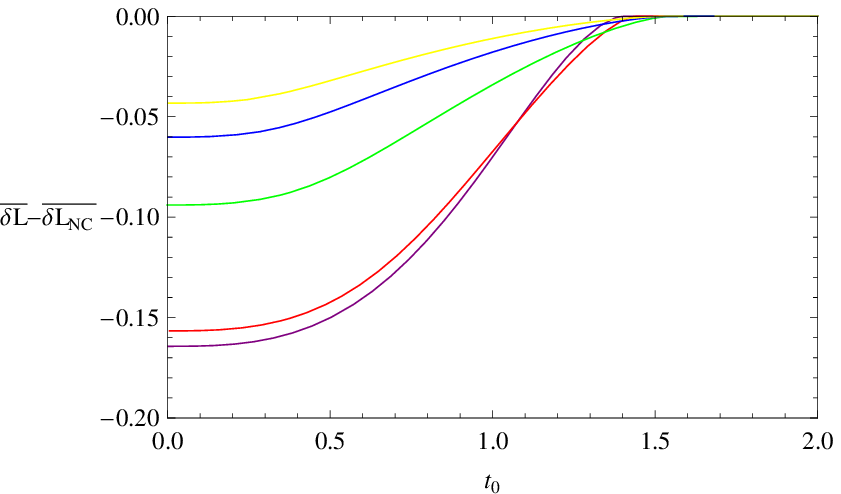} }
\subfigure[$l=3.6$]{
\includegraphics[scale=0.55]{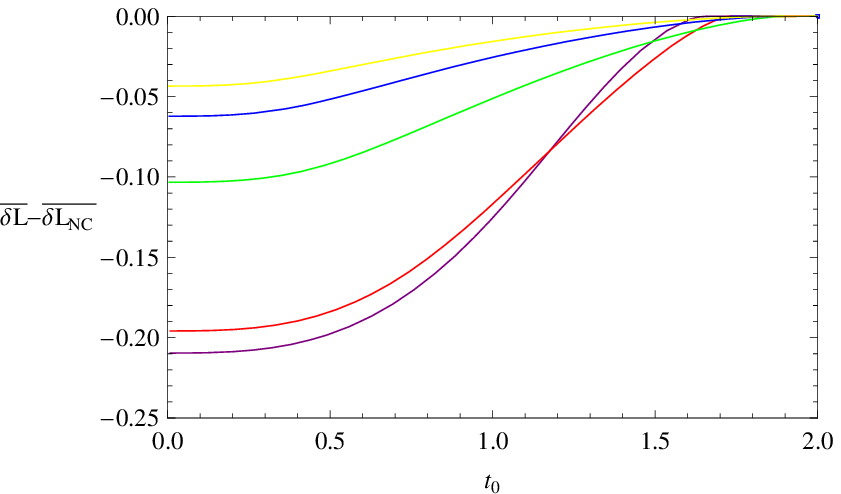}
}
 \caption{\small Thermalization of the renormalized geodesic lengths  for different $\theta$ at a fixed boundary separation.
 The yellow line, blue line,  green line, red line and purple line correspond to  $\theta=0.7, 0.5, 0.3, 0.1, 0.01$  respectively.} \label{fig4}
\end{figure}

With the numerical result of $z(x)$, we can study
the  renormalized geodesic length. As done in \cite{Balasubramanian1, Balasubramanian2, GS},
we are interested in the $l$ independent quantity $\overline{\delta L}-\overline{\delta L_{NC}}$
with  $\overline{\delta L_{NC}}\equiv  \delta L_{NC}/l$ being the  length of the late stage.
Figure (\ref{fig4}) gives the relation between the  renormalized geodesic length and thermalization time for different noncommutative parameters $\theta$ at a fixed boundary separation. From Figure (\ref{fig4}), we know that
for large noncommutative parameters, $\theta=0.3, 0.5, 0.7$, though the background spacetimes have not horizons, the shell will not collapse all the time. At the last stage, they will stop in a stable state at the same thermalization time. But for different $\theta$, the thermalization velocity is different, which can be read off from the slope of the thermalization curve. It should be noted that the thermalization time for the background spacetime without a horizon, $\theta=0.3, 0.5, 0.7$,  is longer than that with a horizon, $\theta=0.01, 0.1$. That is, as a static black brane is formed the shell for large $\theta$ is still collapsing. In the small $\theta$ region, we can observe that the thermalization time
increases as
$\theta$ raises. As the  boundary separation raises, this effect is more obvious, please see (a) and (b) in Figure (\ref{fig4}).   Therefore we know that as the noncommutative parameter increases, the thermalization will be delayed.
 This phenomenon has been also observed previously when we study the motion profile of the geodesic.
 In addition, in Figure (\ref{fig4}), we find for a fixed boundary separation there is always a time range in which  the  renormalized geodesic length for different $\theta$ takes the same value nearly. That is, during that time range,
the noncommutative parameters have few effect on the   renormalized geodesic length. In \cite{Zeng2013, Zeng2014}, effect of the  Gauss-Bonnet coefficient on the thermalization time was investigated, they  also found this phenomenon.

\begin{figure}
\centering
\subfigure{
\includegraphics[scale=0.65]{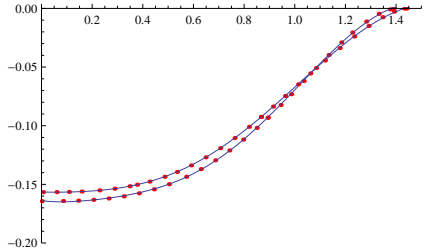}}
 \caption{\small Comparison of the function in Eq.(3.10) with the numerical result for the case $\theta=0.01, 0.1$ at the boundary separation $l=3$. } \label{fig5}
\end{figure}

\begin{figure}
\centering
\subfigure[$\texttt{Thermalization velocity}$]{
\includegraphics[scale=0.23]{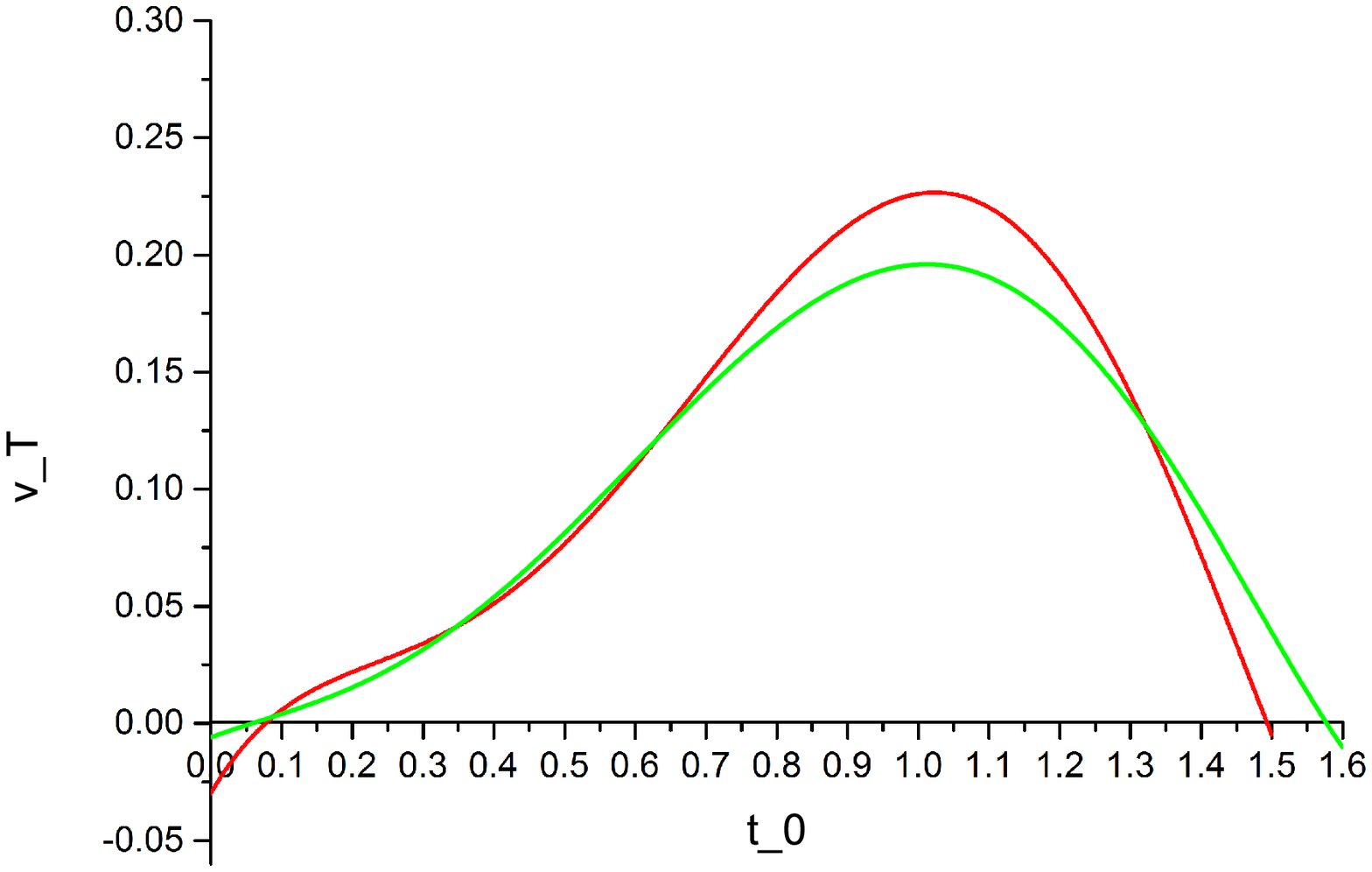} }
\subfigure[$\texttt{Thermalization acceleration}$]{
\includegraphics[scale=0.2]{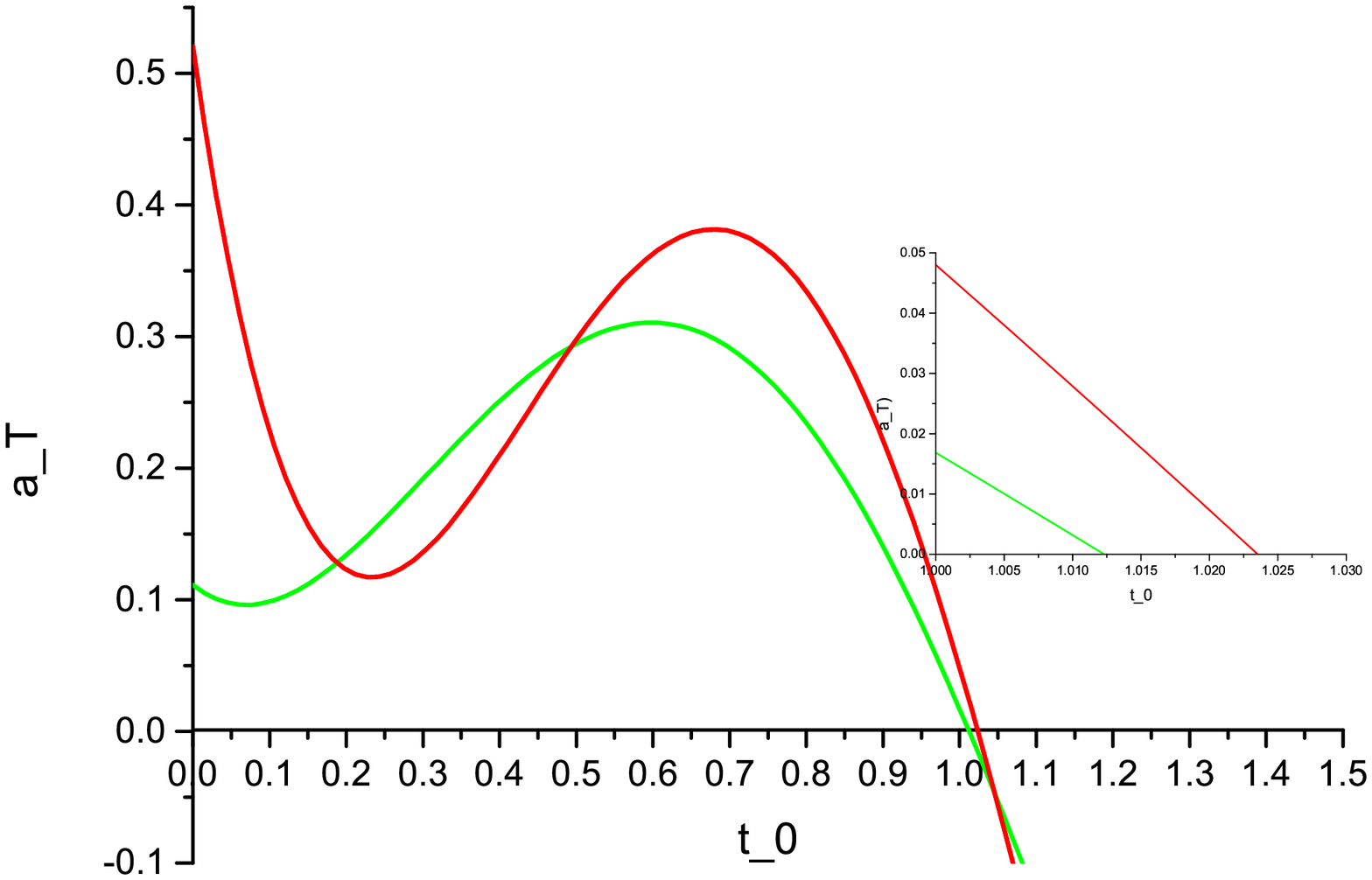} }
 \caption{\small Thermalization velocity and acceleration of the renormalized geodesic  in a noncommutative Vaidya AdS  black brane. The red line and green line correspond to $\theta=0.01$ and $\theta=0.1$.} \label{fig66}
\end{figure}

Interestingly, we find the thermalization curve for a fixed noncommutative  parameter in Figure (\ref{fig4})
 can be fitted as a function of $t_0$. Here we take the case  $\theta=0.01, 0.1$ as examples.  At the boundary separation $l=3$, the numeric curves for  $\theta=0.01, 0.1$ can be fitted as
\begin{eqnarray}
\begin{cases}
g_{0.01}=-0.163841-0.0300155 t_0+0.260142 t_0^2-0.685112 t_0^3\\
  ~~~~~~~~~ +1.10426 t_0^4-0.703409 t_0^5+0.148517 t_0^6\\
g_{0.1}=-0.156499-0.00586248 t_0+0.0557009 t_0^2-0.0778527 t_0^3\\
  ~~~~~~~~~ +0.32609 t_0^4-0.272237 t_0^5+0.0634748 t_0^6
\end{cases}
\end{eqnarray}
 Figure (\ref{fig5}) is the comparison result of the numerical  curves and fitting function curves. It is obvious that at
the order of  $t_0^6$, the thermalization curve can be described well by the fitting function
\footnote{For higher order power of $t_0$, we find it has few contributions
    to the thermalization, including the phase transition point which will be discussed next. }.
   With this function, we can get the thermalization
velocity, defined by $v_{-}T\equiv d(\overline{\delta L}-\overline{\delta L_{NC}})/dt$,  and thermalization acceleration,  defined by $a_{-}T\equiv d^2(\overline{\delta L}-\overline{\delta L_{NC}})/dt^2$, which are plotted in   Figure (\ref{fig66}).
 From the velocity curve, we can observe
 that there is a phase transition point at the middle stage of the thermalization, which divides the thermalization into an
accelerating and a decelerating phase. The phase transition points for different noncommutative parameters can be read off from the null point of the
acceleration curve.
It is easy to find that in the time range,
$0<t_0<1.024$ for $\theta=0.01$ and $0<t_0<1.0125$ for $\theta=0.1$, the thermalization is an accelerating process while for the other time range,  it is a decelerating
 process before it approaches to the equilibrium state. Obviously, as the noncommutative parameter increases, the value of the phase transition point decreases. That is, larger the noncommutative parameter is, earlier the thermalization decelerates. This result  also indicates that the large   noncommutative parameter  delays the thermalization.

From the acceleration curve, we find that  during the acceleration phase, the acceleration is not enhanced always, which first decreases, then increases, and decreases once again. In other words, the acceleration undergos two phase transition during the thermalization.

\section{Conclusions}
Gravitational collapse of a thin shell in the noncommutative geometry is probed by the  renormalized geodesic length, which is dual to probe the thermalization in conformal field theory by the two-point functions. We  first study the motion profiles of the geodesic,   and then   the  renormalized
geodesic length.   For the spacetime without a horizon, we find the shell will not collapse all the time but will stop in a stable state at the same thermalization time. For the spacetime with a horizon,
we investigate how the noncommutative parameter affects the thermalization process by numerical calculation and fitting function.
From the numerical results, we know  that the noncommutative parameter delays
   the thermalization process. In \cite{GS,CK, Yang1, Zeng2014}, effect of the charge on the thermalization time is investigated. They found that as the charge increases, the  thermalization time decreases. Obviously, the noncommutative parameter has the similar effect on the thermalization time as the charge\footnote{In this case, how to distinguish the effect of charge and noncommutative parameters on the thermalization time becomes necessary and important, we  will  address this problem later. }. In addition, for both the thermalization probes,
 we observe an overlapped region where the
noncommutative parameter has few influence on them for a fixed boundary separation. In fact, this phenomenon has also been observed in modified gravity\cite{Zeng2013, Zeng2014}. It is explained that this effect arises from the difference of the temperature of
 the dual conformal field for the thermalization only becomes fully
apparent at distances of the order of the thermal screening length $\tilde{l}_D\sim(\pi T)^{-1}$, where $T$ is the temperature of
 the dual conformal field.

We also find the fitting functions of the thermalization curves. With it, we
get some useful information about the thermalization. We first get the thermalization velocity  at a  fixed
noncommutative  parameter.  From the velocity curve,  we know that
the thermalization is non-monotonic, which is indicated by the negative value of the thermalization velocity at the initial thermalization time.  Secondly  we find there is a
phase transition  point during the thermalization, which  divides the thermalization  into an acceleration phase and
a deceleration phase.  The phase transition point is found to be decreased as the noncommutative parameter increases.
We also obtain the thermalization acceleration, which  is found to be not enhanced always during the acceleration phase.
Recently Liu et al. \cite{liu1, liu2}, followed by \cite{follow},  have investigated the nonlocal observables analytically. They found that the thermalization can be divided into four regimes:  pre-local-equilibration quadratic
growth  regime, post-local-equilibration linear growth  regime, a late-time regime, and a saturation regime. In each regime, they obtained the analytical functions of the nonlocal observables, which are shown to be the linear function of the thermalization time. Obviously, our result agrees with their result in part for we also obtain this linear relation.


\section*{Acknowledgements}
We are grateful to Hongbao Zhang for   his  various valuable suggestions about this work.
 This work is supported  by the National
 Natural Science Foundation of China (Grant Nos. 11405016, 11365008, 61364030) and the natural science foundation of Hubei Province (No. 2014CFB608).
.








\end{document}